\documentclass[useAMS,usenatbib]{mn2e}
\usepackage{graphicx}
\usepackage{rotating} 
\usepackage{longtable}
\usepackage{lscape}
\usepackage[usenames]{color}

\def\lsim{\mathrel{\rlap{\lower 3pt \hbox{$\sim$}} \raise 2.0pt \hbox{$<$}}}
\def\gsim{\mathrel{\rlap{\lower 3pt \hbox{$\sim$}} \raise 2.0pt \hbox{$>$}}}
\def\nat{Nat}
\def\apj{ApJ}
\def\apjl{ApJL}
\def\apjs{ApJS}
\def\mnras{MNRAS}

\def\pasp{PASP}

\def\prd{PhRvD}
\def\kms{\rm km\,s^{-1}} 
\def\msun{\rm M_{\odot}}

\title[Line flux ratio diagnosis for sub-pc binary black holes]{Search of sub-parsec massive binary black holes through
line diagnosis}

\author[Montuori et al.]{C. Montuori,$^{1}$\thanks{carmen.montuori@uninsubria.it} M. Dotti,$^{2}$  M. Colpi,$^{3}$ R. Decarli,$^{4}$ F. Haardt$^{1}$\\
             $^1$ Universit\`a degli Studi dell'Insubria, Via Valleggio 11,
             22100 Como, Italy\\
             $^2$ Max Planck Institut f\"{u}r Astrophysik, Karl Schwarzschild
             Str. 1, 85748 Garching, Germany\\
             $^3$ Universit\`a degli Studi di Milano-Bicocca, Piazza della
             Scienza 3, 20126 Milano, Italy\\
             $^4$ Max Planck Institut f\"{u}r Astronomie, K\"{o}nigstuhl 17,
             69117 Heidelberg, Germany}
\pagerange{\pageref{firstpage}--\pageref{lastpage}} \pubyear{2010}

\begin{document}        
\maketitle
\label{firstpage}

\begin{abstract}
We investigate on the spectral properties of an active black hole,
member of a massive ($10^7-10^9\, \msun$) sub-parsec black hole binary.   
We work under the hypothesis that the binary, surrounded by a circum-binary disc, 
has cleared a gap, and that accretion occurs onto the secondary black hole fed by material closer to the inner edge of the
disc. Broad line emission clouds orbit around the active black hole and
suffer erosion due to tidal truncation at the Roche Lobe surface, following gap opening and
orbital decay. We consider three of the most prominent broad emission lines observed in the spectra of
AGNs, i.e. CIV, MgII and $\rm{H\beta}$, and compute  the flux ratios between
the lines of MgII and CIV ($F_{\rm MgII}/F_{\rm CIV}$) and those of MgII and
${\rm H{\beta}}$ ($F_{\rm MgII}/F_{\rm H{\beta}}$). 
We find that close black hole binaries have $F_{\rm MgII}/F_{\rm CIV}$ up to one order of magnitude
smaller than single black holes. By contrast $F_{\rm MgII}/F_{\rm H{\beta}}$ may be significantly reduced 
only at the shortest separations. Peculiarly low values of line flux ratios together with 
large velocity offsets between the broad and narrow emission lines and/or periodic variability  
in the continuum (on timescales $\gsim$ years) would identify genuine sub-pc binary candidates.


\end{abstract}

\section{Introduction}\label{sec:intro}

Massive black hole {\it binaries} are considered to be a direct outcome of
galaxy mergers. They form in the advanced stages of
the galactic interaction as soon as the mass of the two 
black holes (BHs) exceeds the gas/stellar mass enclosed
within their orbit, on scales of $\sim 1-10$ pc
(Begelman, Blandford \& Rees 1980; Merritt \& Milosavljevic
2005; Mayer et al. 2007; Colpi \& Dotti 2009). 
In gas rich environments and in the advanced stages
of BH hardening, the binary is likely to be surrounded by a circum-binary accretion disc.
In the interaction with the disc 
the binary transfers orbital angular momentum through gravitational  torques
after having excavated a gap, i.e. a hollow density region (e.g. Artymovicz \& Lubow 1994).
Migration of the secondary BH toward the primary is then regulated by the
rate at which viscous torques in the disc respond to the tidal field of the binary.
If migration continues to be effective the BH binary is expected to enter the gravitational wave (GW) 
driven regime, and the transit to this state is expected to be the
longest-lived one (e.g. Ivanov et al. 1999; Gould \& Rix 2000; Milosavljevic
\& Phinney 2005; Armitage \& Natarajan 2002; MacFadyen \&
Milosavljevic 2008; Lodato et al. 2009; Cuadra et al. 2009; Loeb 2010).
The observability of BH binaries with separations $\lsim 0.1$ pc, 
at the stage of disc/GW driven migration, is the focus of our study.

Sub-parsec binary AGN, if the BHs are active, can be viewed as key tracers of 
galactic mergers along the co-evolution of BHs and spheroids (Ferrarese \& Ford 2005).
Direct imaging of binary AGN on parsec scales however can be
carried on, in the near Universe, only in the radio band, when both BHs are active, thanks
to the use of radio interferometers such as the very long baseline array
(VLBA). This technique needs to point precisely at the source whose position in
the sky should be known in advance. So far, this method yielded one candidate, 0402+379, 
a source with two compact radio cores seen at a projected separation of $\sim 7$ pc, 
representing  the closest system imaged to date (Rodriguez et al. 2006).

Spectroscopic studies may provide an alternative method for
revealing binary AGN, at even closer separations, 
through the search of velocity shifts in the multiple line systems 
resulting from the Keplerian motion of the two BHs. 

Peculiar spectra with large velocity shifts of the order of
$\sim 10^3\,\kms$ are easily selected among thousands of
quasar spectra, and five spectroscopic BH binary candidates have been found so far in the public
archives of SDSS survey: J092712.65+294344.0 (Bogdanovic et al. 2009; 
Dotti et al. 2009), J153636.22+044127.0 (Boroson \& Lauer 2009),
J105041.35+345631.3 (Shields et al. 2009), J100021.80+223318.6, also 
known as 4C+22.25 (Decarli et al. 2010), and J093201.60+031858.7 (Barrows et
al. 2010). All these sources show shifted line systems, but their spectra differ significantly from one another and
the interpretation of their physical nature is not unique 
(Eracleous et al. 2004; Komossa et al. 2008; Shields et al. 2009, 2009b; Heckman et
al. 2009; Wrobel \& Laor 2009; Decarli et al. 2009, 2009b; Chornock et
al. 2010; Barrows et al. 2010).
We further note that despite it's simplicity this method is biased in
favour of binary systems whose orbital configuration maximizes the velocity shifts between different
line systems. Moreover it can not be applied to search for BHB candidates at redshift
$z\gsim 2$ since the rest-frame wavelength of the main narrow emission
lines is $\gsim 3000$ \AA{}. 

In light of these findings, is it still possible to identify other complementary
signatures of a binary system at the stage of viscous-driven migration or GW-inspiral? 

High resolution numerical simulations can
play an important role in predicting the electromagnetic signatures
associated with binary systems. Nevertheless hydrodynamical simulations with 
full radiative transfer are still out of reach due to 
the very high computational cost. As illustrated in Bogdanovic et al. 
(2008; 2009b), the use of photoionization codes in
combination with hydrodynamical simulations allows to analyse the
observational features of a sub-pc binary system, interacting with a
gaseous disc, just for few orbital periods.  
Relying on a different approach,  
Shen \& Loeb (2009) studied the potential signature left in the spectra 
by two active BHs surrounded by BEL clouds that initially are
inside the hole's Hill spheres. 
The spectra display double peaked features 
which over years show coherent radial velocity drifts due to the 
Keplerian motion. However, since BELs carry individually 
large widths, only in a restricted range of binary separations 
BELs display line-of-sight orbital velocity differences larger than that of their intrinsic FWHM. 
Furthermore, line profiles become complex as soon as the two BLRs 
start to move in the combined gravitational potential of the BH binary.

Here we devise a new approach to search for BHB candidates starting from
single-epoch, optical spectroscopy datasets, relying on the possibility that
the presence of a binary system can affect the flux ratios between broad
emission lines. We explore the case where one 
BH (the lighter secondary) is active after the BH binary has cleared a
gap inside the circum-binary disc, and compute line flux ratios
emerging from a BEL system anchored to the secondary BH. This configuration 
differs from that considered in Bogdanovic et al. (2008) and in Shen \&
Loeb (2009) and is motivated by the results of previous numerical studies on the 
evolution of black hole binary systems within gaseous environment. 

The outline of the paper is as follows: 
in Section \ref{sec:model} we briefly describe
the accretion and BH binary evolution model. In Section \ref{sec:method} 
we describe how we model the BLR and compute the flux ratios of different lines using  
the photoionization code CLOUDY. In Section  \ref{sec:results} we present our results.
The possible use of these results as tool in the search of new secure spectroscopic
BHB candidates is described in Section \ref{sec:search}. In Section \ref{sec:discussion} 
are our discussion and concluding remarks.

\section{Binary black holes in circum-binary  accretion discs} \label{sec:model}

Consider a circular BH binary with semimajor axis $a$ and   
mass ratio $q\equiv M_2/M_1$ between the secondary (${M_2}$) and primary (${M_1}$) BH, and 
let $q=0.3$, in accordance with recent findings that indicate the formation of close BH pairs 
(and so of binaries) only in mergers with sufficiently large
mass ratios to avoid premature disruption of the less massive galaxy by tides and ram-pressure
stripping (Kazantzidis et al. 2005;  Callegari et al. 2009).  

The BH binary is assumed to be embedded in a circum-binary geometrically thin accretion disc coplanar with the
orbital plane of the BH binary (Ivanov et al. 1999). 
For $q\gsim 0.1$, the binary strongly perturbs the disc 
by exchanging angular momentum with the gas and excavates a gap, i.e. a cavity extending up to $ r \sim 2a$, where $a$ is the orbital separation (e.g. Artymovicz \& Lubow 1994). Following gap opening, the primary, left in a state
of low accretion, consumes most of its disc over the inner viscous time scale.
The secondary, instead, that moves closer to the inner rim of the
circum-binary disc, has lower 
speed relative to the gas, and continues to accrete higher density 
material growing its own disc (e.g. Hayasaki et al. 2007, 2008; Cuadra et al. 2009). 

Here we work under the hypothesis that  the lighter secondary
is the only active member of the binary, and that a BEL system is maintained 
around the secondary BH, through binary evolution. 
BEL-clouds  orbit inside the Roche lobe of
$M_2,$ and their localized emission represents here the main contribution 
to the emitted spectrum as it originates in 
the high density region of the inner disc bound to $M_2$.
Due to the orbital decay induced by viscous torques or GW emission, the Roche radius 
${R_{\rm L}}\sim 0.49 \, a \, q^{2/3}/[0.6 q^{2/3}+\ln(1+q^{1/3})] $  (Eggleton 1983) 
decreases  
and we expect a progressive erosion of the BLR: the BLR gas, no longer bound
to the secondary BH, experiences, outside $R_{\rm L}$, the tidal torque from the binary and is dragged away.
We thus work under the hypothesis that the BLR is tidally truncated at $R_{\rm L}$
as the binary orbit shrinks under the action of external torques.

In the interpretation of the results it is important to define a particular
phase, referred to as binary-disc decoupling, that occurs at the bridge between the viscous and the GW driven
domains. As described in Haiman, Kocsis \& Menou (2009), there exists a time
in which the inner edge of the circum-binary thin disc can no longer follow the
migration of the secondary as the inward viscous diffusion timescale in the
disc becomes longer than the timescale of GW-driven
inspiral ($t_{\rm{GW}}$).  At such a separation the circum-binary disc decouples
and the secondary might be no longer fueled. In this case, from the decoupling
time on, the secondary active BH is expected to consume its disc on a timescale
that is $\gsim t_{\rm{GW}}$ (see eq. 21 of Haiman et al. 2009 for an expression
of the viscous time in the case of a steady-state thin accretion disc). 
Thus we can assume that the activity of
the secondary BH can be sustained throughout the binary orbital decay.

The question we want to address is the following: is there an observable  signature in the 
emitted spectrum associated to this evolutionary scenario? 
We expect that the emission from the peripheral BELs becomes inefficient below
some critical orbital separation.  This can be especially relevant for 
low-ionization lines, such as that of MgII and $\rm{H\beta}$, in comparison
to high-ionization ones, since the former are emitted preferentially at
greater distances from the source of the ionizing flux.  To this purpose,
in the next Section we devise a method to compute the expected effects of
the BLR tidal truncation on the flux ratios.

\section{Flux ratios}\label{sec:method}

We consider three of the most prominent BLR lines observed in the spectra of
AGNs: CIV, MgII and $\rm{H\beta}$. We then  
focus on the flux ratios between those lines that can be simultaneously
observed in an optical survey, such as the SDSS, up to
redshift $z \sim 2$.
In particular we compute the flux ratios between the lines of MgII and CIV 
($F_{\rm MgII}/F_{\rm CIV}$) and those of MgII and 
${\rm H{\beta}}$ ($F_{\rm MgII}/F_{\rm H{\beta}}$). 
The flux ratios are calculated for different sizes of the BLR around the secondary
according to the scenario outlined above. For a circular binary of mass ratio $q$, secondary mass $M_2$ and separation $a$, the BLR is  truncated at the Roche radius $R_{\rm L}$.
The secondary is assumed to emit a luminosity $L=f_{\rm Edd} L_{\rm Edd}$
with a constant Eddington factor $f_{\rm Edd}$.
The greatest orbital separation $a_{\rm o}$ is set under the assumption that 
$R_{\rm L}(a_{\rm o})$ coincides with the size $R_{\rm BLR}$ of a BLR of an isolated BH
accreting at $f_{\rm Edd}$ as expected from the observational relation of $R_{\rm BLR}$ with
the luminosity at 5100 \AA{} (Kaspi et al. 2005), considering $\lambda L_{\lambda}
    (5100) \sim (1/9) \, L$ (Kaspi et al. 2000). 

In order to compute the flux ratios at each separation, i.e. 
at each orbital period $P(a)$, we use  the photoionization code CLOUDY (version 08.00; Ferland et al. 1998). To map the BELs we refer  to the ``locally optimally cloud'' model
(LOC; e.g. Baldwin et al. 1995).\footnote {Considering current uncertainties on the nature and geometry of the BLR,
we choose this model  compared to more complex ones (see review of Gaskell 2009 and references therein).}
Following Korista et al. (1997) and Korista \& Goad (2000), we compute a
grid of photoionization models assuming each cloud as a slab of
constant gas density with solar metallicity and a clear view to the
ionizing flux. The shape of the ionizing continuum is
taken as one of the templates for a radio-quiet active galaxy stored 
as part of the CLOUDY code.\footnote{In particular we set the incident continuum
with the ``table AGN'' command.} 
The
column density is set to $N_{\rm H}=10^{23} \, {\rm cm^{-2}}$ and is kept
fixed for all clouds.  
 We consider the contribution of clouds with $8\le \log \, (n_{\rm H}/{\rm cm^{-3}})
  \le 14$ and hydrogen ionizing flux ${18}\le \log\, (\Phi_{\rm H}/{\rm cm^{-2}\,sec^{-1}}) \le 24.$\footnote {These
limits are motivated by physical or observational considerations: clouds located
further away from the ionizing source (log $({\Phi_{\rm H}/{\rm cm^{-2}\,sec^{-1}}) < 18}$) give a very
low contribution to the emission and form graphite grains, while the absence of
broad forbidden lines implies that clouds with lower density are not
present.  }
According to the LOC model the main properties of quasar spectra are 
dominated by selection effects of atomic physics and radiative transfer that determine,
for each individual BLR cloud, 
the reprocessing efficiency of the ionizing continuum into line radiation. In this contest,
it is assumed that there exists a spread in gas properties at each radius in the BLR. 
Therefore in order to compute the total BELs flux, we have to consider the contribution 
of each cloud lying in the density-flux plane spanned by the photoionization grid models.
In particular it was shown that one can sum over all these contributions by making 
the simplifying assumption that clouds are distributed in gas density and distance 
following a weighting function separable in both variables (e.g. Baldwin et al. 1997).
\footnote {These assumptions usually refer to a spherically symmetric
distribution of BLR clouds but they are still consistent with the case of a
thick disc, which is the expected distribution for the BLR gas in the model described in section \ref{sec:model}. 
Both geometries imply a high covering factor which is in
agreement with the observations (e.g. Gaskell 2009). On the other hand a 
high covering factor and a disc like geometry imply that the effect of cloud self-shielding 
may become relevant. In this case it has been shown that regions emitting BELs of different 
ionization potential are more clearly spatially separated (Gaskell 2009 and references therein). 
We expect that accounting for BLR self-shielding the effect of the  
the BLR erosion on the $F_{\rm MgII}/F_{\rm CIV}$ ratio would be even stronger than what obtained in our calculations.} 
  
In our study, we consider the case of a uniform
distribution of cloud distances and densities, and a second case 
in which the weighting function is a power law with index -1
in both variables (power-law model, here on). 
We set the density range for the two different distributions 
such that at the largest orbital separation the computed flux ratios
are consistent with the typical values observed in the AGN spectra, in particular: $
9 \le \log \, (n_{\rm H}/{\rm cm^{-3}}) \le 13$ for the homogeneous model, and $9 \le \log \, (n_{\rm H}/{\rm cm^{-3}}) \le 14$
for the power-law model. Considering the adopted cloud distributions 
in density and space, the resulting mean number density as function of distance are:
$n_{\rm{H}} \sim 5 \times 10^{12} \, \rm{cm^{-3}}$, constant with radius for
the uniform BLR model, and $n_{\rm{H}} \sim 8 \times 10^{12} \,
(r_{\rm{in}}/r) \,\, \rm{cm^{-3}}$  for the
power law case, where $r_{\rm{in}}$ is the inner BLR radius.
In the power law case the contribution from higher density gas at greater distances is
less relevant than in the uniform model. The need of a higher density limit in the power 
law model to reproduce the observed values of the $F_{\rm MgII}/F_{\rm CIV}$
ratio can be therefore understood considering that the MgII line is more
efficiently reprocessed at higher densities and lower fluxes than that of CIV.

\section{Results} \label{sec:results}
 \begin{figure}
\centering
\includegraphics [scale=0.45] {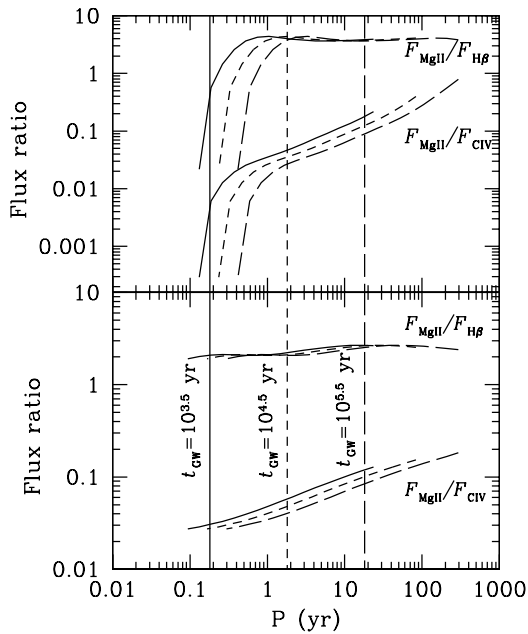}
\caption{\small{Flux ratios between the broad
    emission lines of MgII and CIV ($F_{\rm MgII}/F_{\rm CIV}$) and those of MgII and $\rm{H\beta}$
    ($F_{\rm MgII}/F_{\rm H\beta}$) as function of the
    orbital period $P$ for a BH binary with
    different total masses. The mass ratio is set to $q=0.3$ and the Eddington factor
    to $f_{\rm Edd}=0.1$. Solid/dashed/long-dashed lines refer to a secondary mass of
    $M_2=10^7-10^8-10^9 \, \msun$, respectively. The vertical lines mark the orbital period 
    at the time the binary 
    detaches from the circum-binary disc. Upper (Bottom) panel corresponds to the
    uniform cloud (power-law) model. }}
\label{fig:period}
\end{figure}

  \begin{figure}
\centering
\includegraphics [scale=0.45] {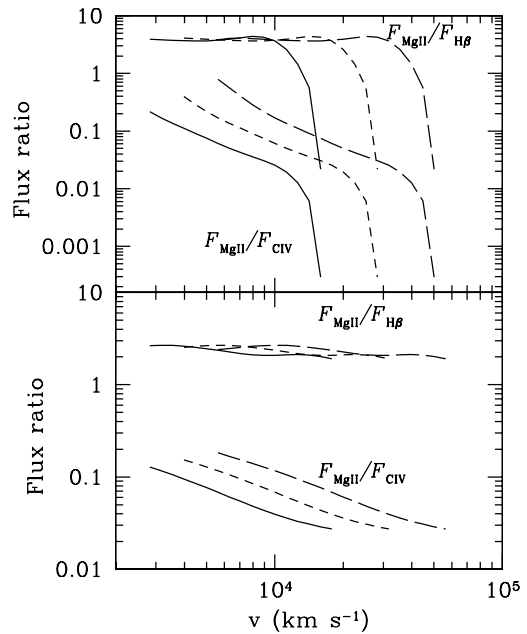}
\caption{\small{Line flux ratio (as in Figure \ref{fig:period}) as a function
of the secondary BH orbital velocity relative to the center of mass. The velocity reported here would
express the velocity offset between the BEL and the NEL systems for a BH binary seen 
   edge-on and at one of the two orbital nodes.}}
\label{fig:velocity}
\end{figure}

\begin{figure}
\centering
\includegraphics [scale=0.45] {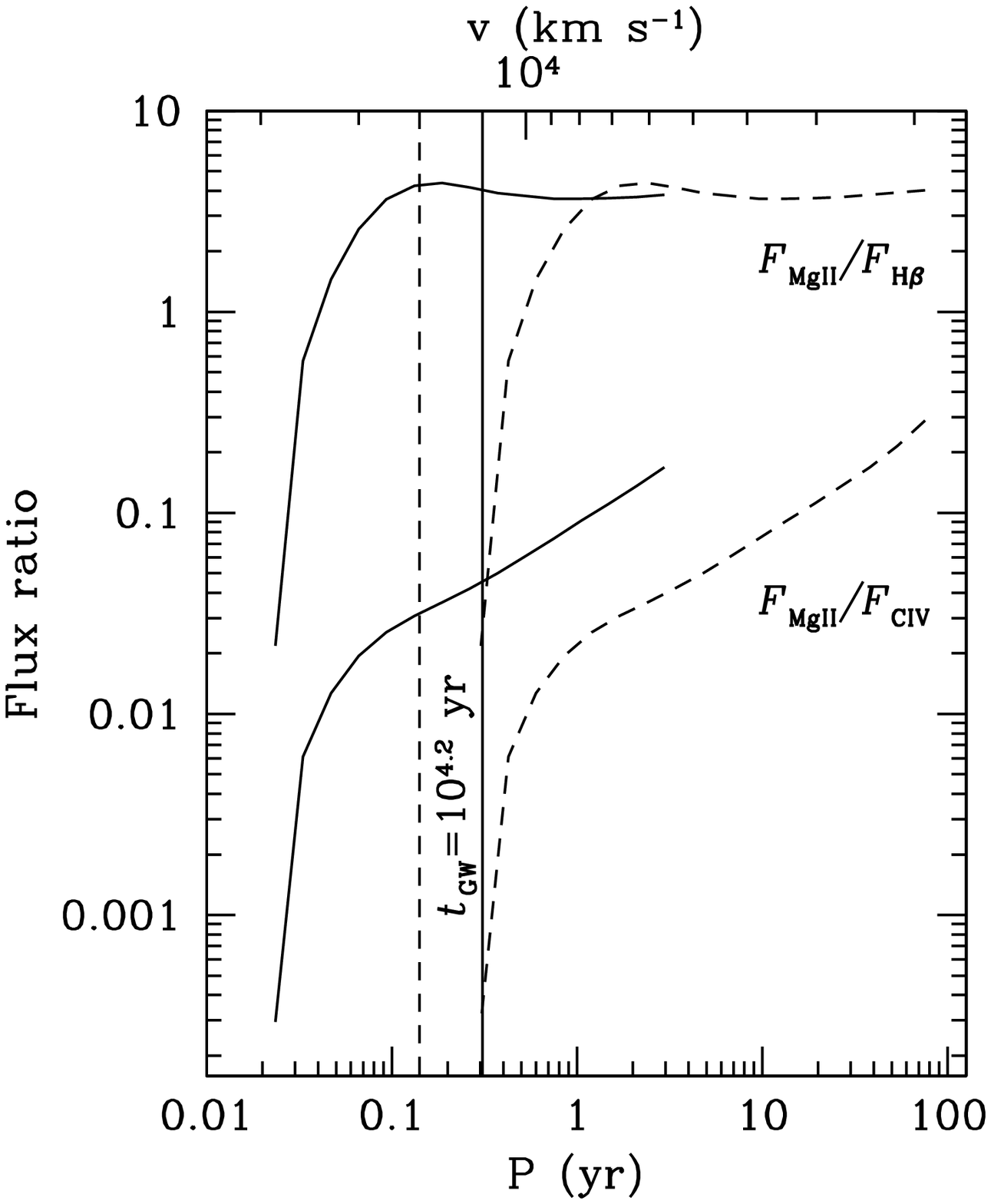}
\caption{\small{Line flux ratios 
for two different values for $f_{\rm Edd}$, for the 
uniform cloud model with $q=0.3$ and $M_2=10^7 \,\msun. $
Solid (dashed) lines refer to $f_{\rm Edd}=0.01$ (0.3). 
The line ratio curves shift to shorter periods for lower values of 
$f_{\rm Edd}$ while the opposite occurs for higher Eddington factors. 
As explained in the text, this is due to the correlation between 
the unperturbed outer radius of the secondary BLR and it's luminosity. 
Lower  values of $f_{\rm Edd}$  imply longer 
periods in correspondence of binary-disc 
decoupling because of the longer viscous time scale. }}
\label{fig:fedd}
\end{figure}

  Figure \ref{fig:period} shows the flux ratios $F_{\rm MgII}/F_{\rm CIV}$ and  $F_{\rm MgII}/F_{\rm H\beta}$ 
  as a function of the Keplerian orbital period of the BH binary, for $q=0.3$
  and for a secondary with $M_2=10^7,10^8,10^9 \msun.$
  The upper panel refers to the case of uniformly distributed BLR clouds,
  while the lower to the power-law model.
    
  We find that the ratio $F_{\rm MgII}/F_{\rm CIV}$ is a decreasing function of the orbital separation,
  and so of $P$, and therefore it can be regarded as a 
  possible diagnostic tool in the spectroscopic search of binary systems (see Sec. \ref{sec:search}). On the other hand  
  the $F_{\rm MgII}/F_{\rm H\beta}$ ratio is fairly constant 
  apart from a sharp decline present only in the case of a uniform 
  clouds distribution, that will be discussed below. Therefore $F_{\rm MgII}/F_{\rm H\beta}$ 
  is not sensitive to the truncation of the BLR while $F_{\rm MgII}/F_{\rm CIV}$   
  shows lower values for closer separations in both BLR models and is already reduced of an order of magnitude
  before the drop that occurs in the uniform distribution case.
  These results are consistent with what expected considering that lines of different ionization potential, such as those
  of CIV and MgII, are preferentially emitted in spatially separated regions. 
  As mentioned before, in the case of a uniform BLR, both flux ratios decrease rapidly around 
  a critical orbital period $P_{\rm drop}$. This rapid decay is not observed for the
  case of a power-law BLR distribution. 
  As described in Section 3, observations require that the BLR cloud density does not exceed
  $\log \, (n_{\rm H}/{\rm cm^{-3}})\lsim 13$ for the uniform model.
  Higher densities needed in the power-law model can still contribute to the MgII line emission
  at closer orbital separations, preventing the occurrence of the drop in the line flux ratio.
  However  we expect to observe at most a few binary systems 
  at such small periods ($P < P_{\rm drop}$) because of the short coalescing timescale for GW emission.

  Vertical lines refer to the orbital period $P_{\rm{dec}}$, for a selected binary, below which
  the binary decouples from the circum-binary disc. As described in Section \ref{sec:model}, at this stage the BH binary is already in the 
  GW-driven inspiral regime and in Figure 1 we report  the corresponding 
  times for GW-driven coalescence, in the range $3\times 10^3-10^5 \,{\rm  yr}$. This critical period is 
  computed following Haiman et al. (2009) for a steady geometrically thin accretion disc (see
  their equations 30). We notice that for the parameters chosen in our calculation the binary-disc 
  decoupling occurs at $P_{\rm{dec}}\gsim P_{\rm drop}$ where there is more chance to observe binary systems
  and the values of the $F_{\rm MgII}/F_{\rm CIV}$ ratio can be already reduced up to an order of magnitude, as discussed before.
    
  Figure \ref{fig:velocity} carries the same information of Figure \ref{fig:period} but expresses
  the decrease in the flux ratios as a function of the velocity of the secondary BH relative
  to the center of mass of the binary. The velocity is in principle measurable as
  Doppler shift between the BELs and the Narrow Emission Lines (NELs) that are
  emitted from gas at much larger distances and that provide the rest-frame
  redshift of the host galaxy.  The velocity reported in Figure \ref{fig:velocity} 
  refers to the maximum observable velocity offset, corresponding to a binary
  seen edge-on and at at one of the two orbital nodes, 
  and can exceed $\gsim 10^4\,\kms$ in the region of interest.  

  Figure \ref{fig:fedd} illustrates the effect of varying the Eddington
  factor for the case of a secondary  with $\rm{M_2=10^7 \msun}$, $q=0.3$
  and for a uniform BLR. The flux ratios curves shift to shorter periods for
  lower Eddington factors. The unperturbed outer radius of the BLR of the
  secondary is proportional to its luminosity through the $R_{\rm{BLR}}-L$
  relation of Kaspi et al. (2005), as in the case of isolated BHs. This
  implies that the smaller BLR of a secondary accreting at a lower pace
  starts to be tidally perturbed at smaller binary separations.
  On the other hand for lower values of $f_{\rm Edd}$ the binary-disc
  decoupling may occur at greater orbital separations, due to the longer viscous time scale.
  As will be discussed in Section \ref{sec:discussion} this may affect the relevance of 
  the possible contribution to the observed BELs from the circum-binary disc, not included in our work.

\section{Spectroscopic search of binary candidates}\label{sec:search}
 \begin{figure}
\centering
\includegraphics [scale=0.35] {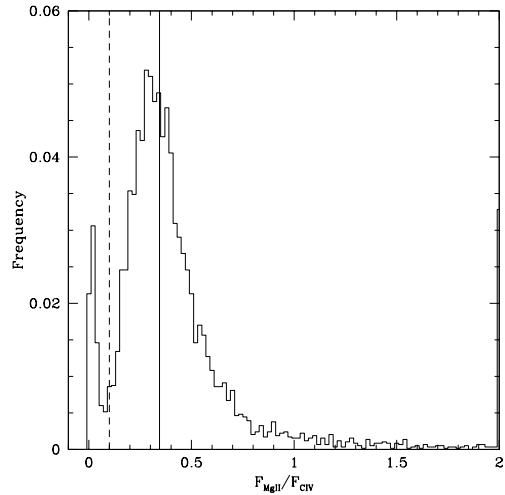}
\caption{\small{Normalized distribution of the flux ratio $F_{\rm MgII}/F_{\rm CIV}$
measured from the SDSS spectra of 5820 quasars at redshift $\sim 2$.
The vertical solid line marks the median value of the flux ratio distribution ( $\sim$ 0.3). 
Sources with flux ratios lower than 0.1 are located on the left of the dashed vertical
line and among them it is possible to look for sub-pc BH binary candidates
according to our results discussed in the text.}}
\label{fig:distribution}
\end{figure}

The huge quantity of data collected by recent surveys
give a unique chance to look for the still elusive observational evidences of close BH binary systems. 
This opportunity leads us to consider the possible implications of our results 
in the spectroscopic search of new BHB candidates among the public archive of the SDSS survey. 

Firstly we notice that the wavelength range covered by the SDSS does not 
allow to simultaneously observe both line ratios considered in our work.
In particular, it is possible to measure the ratio $F_{\rm MgII}/F_{\rm H{\beta}}$ 
for those sources in the redshift bin between $\sim 0.4$ and $\sim 0.8$,
while the ratio $F_{\rm MgII}/F_{\rm CIV}$ targets objects
at redshifts $z \sim 2$. Among the three lines considered here only 
$\rm{H{\beta}}$ is emitted both in the BLR and the NLR. 
Accordingly, measures of $F_{\rm MgII}/F_{\rm H{\beta}}$ with a lower value
than observed for isolated AGN, around redshift $0.4\lsim z \lsim 0.8,$ 
accompanied by the  rapid periodic luminosity variability and velocity offset between
BELs and NELs are in favour of the BH binary hypothesis 
(we refer to Sec. \ref{sec:discussion} for a more detailed discussion  
on these last two spectroscopic signatures). However it has to be taken into 
account that the expected number 
of sub-pc binary systems at low redshift is at most of the order 
of a few (Volonteri, Miller \& Dotti 2009).

On the other hand, at higher redshift, $ z \sim 2$, where
the predicted number of such systems increases, the observation of reduced
$F_{\rm MgII}/F_{\rm CIV}$  can be indicative of the presence of 
a BH binary and can be confirmed again by searching for
luminosity variability on scale $\gsim$ years. 
We further notice that at redshifts $z \ge 2,$ it is difficult
to find evidences of velocity offsets between BELs and NELs since the
wavelength of the most common NELs is longer than $\sim$ 3000 \AA \,(e.g.
Osterbrock \& Ferland 2006). The signature of binary
orbital motion in the AGN spectrum at redshift $\ge 2$ would require
spectroscopic studies in the IR band. Therefore our results can be useful to look for
BHB candidates in a redshift range that has been not yet explored.

Fig. \ref{fig:distribution} represents the normalized distribution of the ratio 
$F_{\rm MgII}/F_{\rm CIV}$ for 5820 quasars at $1.9 \le z \le 2.1$ 
from SDSS DR6 catalog. The median value of the distribution ($\sim 0.3$) 
is marked by the solid vertical line. About 500 objects located on the left of
the dashed vertical line are characterized by a flux ratio value of a factor 
of $\ge 3$ lower that what typically observed. The existence of these
peculiar sources offers a key chance to test the findings of our study. Hence we intend to analyse in a following work (Montuori et al., in
prep) the spectroscopic optical (rest-frame) features of BHB candidates
selected among those objects which in Fig. \ref{fig:distribution} 
show unusually reduced values of $F_{\rm MgII}/F_{\rm CIV}$. 
NIR observations of these selected sources will allow to measure velocity
off-sets between BELs and NELs, in particular between the 
$\rm{H\beta}$ and $\rm{[OIII]_{\lambda 5007}}$ lines.

\section{Discussion \& Conclusions}\label{sec:discussion}
        
According to our model hypothesis, the spectra of a massive BH binary, 
at the orbital stage of viscous/GW driven migration,  
show decreased line flux ratios, compared to what observed for single black holes,
due to the erosion of the BLR at the Roche lobe radius of the active secondary.
In light of our findings, AGN spectra characterized by $F_{\rm MgII}/F_{\rm CIV} \leq 0.1$ 
could be interpreted as a signature of BH binaries
with separations in the range of $\rm{0.001-0.1 \, pc}$, and orbital periods $P$ between $0.1-10$ yr,
depending on the total binary mass in the range of $10^7-10^9\rm{M_{\odot}}$. 
This flux decrease amounts to nearly three orders of magnitude for binary systems at the
shortest separations, in the case of BELs emitted 
by a uniform distribution of cloud properties. For clouds with steeper distributions
in space and density the flux ratios can be diminished of a factor $ \lsim 10.$ 

We note that the flux ratio is particularly low when the 
BHB is in the the short-lived evolutionary stage of GW inspiral.
In this phase, the binary often 
reaches the critical orbital distance below which the circum-binary disc 
decouples from the inspiralling BHs.
This can be important when considering the potential contributions 
to the observed BELs not included in our calculations. 
Depending on the uncertain
physical conditions of the gas in the surroundings of the binary 
(e.g., disc  orientation relative to the binary orbital plane, disc geometry, degree of 
illumination/ionization) the gas located at the inner edge of the 
circum-binary disc may contribute to the BELs, making the predicted
diminished values of the flux ratios 
more difficult to observe. We expect however a weakening of 
the potential emission from clouds residing in the circum-binary disc, should the emission be present, since    
the disc edge freezes and the distance between the emitting BH and this 
gas keeps increasing.

All the results discussed in our work assume binary systems on circular orbits as BH inspiral
in rotating discs leads to circularization of the initial orbit (Dotti et al. 2006; 2007; 2009b). 
After gap opening however, the interaction of the binary with the circumbinary disc 
drives the growth of every small residual eccentricity up to a limiting value of $\approx 0.3$ (Armitage \& Natarajan 2005; Cuadra et
al. 2009). Such a small eccentricity implies a pericenter of the
binary orbit $\approx 0.7$ times smaller than the semimajor axis. If
the disc is tidally truncated at the pericenter and has no time to
re-expand at apocenter (e.g. due to periodic accretion from the circumbinary disc)
the line ratios would be the same as those 
corresponding to a slightly closer binary. The period of the binary
would be at most twice longer than what expected from the line ratios
assuming circular orbits. As a consequence, longer observations would
be required to observe the periodicity signatures in the line 
shifts and in the accretion luminosities discussed hereafter. 

Low values of the line flux ratios
should be accompanied by at least one or two of the following signatures.
The first signature would be the presence of de-projected velocity offsets between
the BELs and the NELs up to $\sim 10^4 \,\kms$. The second would be a change in luminosity over the orbital period $P$.  
Bogdanovic et al. (2008) and Haiman et al. (2009) suggested that the accretion rate on
the active BH can be modulated on the timescale of the binary orbital period.  
The line emission features investigated in our study are occurring at orbital periods 
of the order of months to years. As already mentioned in
Sec. \ref{sec:results}, there is little chance to observe binary systems with
periods $ P \lsim 1 \,$ yr due to their shorter lifetimes. Therefore 
we expect that periodicity both in velocity-offsets and in continuum luminosity can 
be used to verify the binary hypothesis through spectral monitoring of binary systems on
timescales $\geq 1-10 \,$ yr, as it would be feasible for example in the case
of a BHB with $M_2=10^8 \rm{M_{\odot}}$ at $P \sim 15 \,$ yr when $t_{\rm{GW}}\sim 10^7 \,$ yr.
    
Can these features possibly  disentangle genuine binary candidates from the
case of a recoiling BH? The signature of a periodicity in the BELs is not expected 
in the case of a recoiling BH. The BH ejected from the nucleus of the host galaxy  
after binary coalescence would carry away a disc with an outer radius of the order 
of ${R_{\rm out}\sim G M_{\rm BH}/v_{\rm kick}^2}$, where ${M_{\rm BH}}$ and 
${v_{\rm kick}}$ are the mass and the velocity of the recoiling BH, respectively. 
Numerical simulations in general relativity show that the maximum predicted value for the kick velocity is 
$\rm{\lsim 4000 \, km \, s^{-1}}$ (Lousto et al. 2009 and references therein). 
This would correspond to the minimum outer radius for
the disc bound to the ejected BH, $R_{\rm out}$, that we can
compare with the BLR radius, $R_{\rm BLR}$, for an isolated BH,  considering three different values for the mass of the
BH remnant,$M_{\rm BH}=10^7,10^8,10^9 \, \msun$. We
find  $R_{\rm out} /R_{\rm BLR} \sim 1$ for $f_{\rm Edd}\lsim 0.3$. In this case we expect that
the line ratios between the low and the high ionization lines do not
differ from what observed in the case of a standard AGN.
   
This is a first study that aimed at exploring  
the potential effect of the erosion of the BEL system due to orbital motion
around a BH binary with only one BH active (the secondary). 
Our next step will focus on the emission of   
transient streams of matter inflowing from the circum-binary disc on to the 
BHs  (Bogdanovic et al. 2009; Cuadra et al. 2009), the presence of lower density gas in the gap
region (Dotti et al. 2009), and the shape of the emission lines as function of the BLR geometry and dynamics.

As an example we notice that
the binary candidate 4C+22.25, described in the recent work of Decarli et
al. (2010), can represent an ideal case to test and improve the predictions of our
study. The binary model consistent with the observations of this source requires
the presence of a single active BH, which is supposed to be the secondary, a total mass
of $\sim 10^9 \msun$ and an orbital period of the order of $\gsim 30$ yr.
 According to our 
findings, these orbital parameters imply that the BLR of the active BH
is tidally perturbed with respect to the case of an isolated AGN. As already reported in the letter of Decarli et
al. (2010), one of the peculiarities of the optical 
spectrum of 4C+22.25 is that the BELs are very broad and faint. This can be
interpreted as a further observational evidence of the truncation of the secondary
BLR that is present together with the reduced flux ratios. Further observations
of the source, possibly at different wavelengths, will help to constrain
the binary model and in particular the scenario proposed in our work, for example through the
analysis of  BEL profiles and of flux ratios between BELs of different ionization potential.

\section*{ACKNOWLEDGEMENTS}
The authors thank the anonymous Referee for his/her constructive comments that
greatly helped to improve this manuscript.

\label{lastpage}

\end{document}